\documentclass{aa}  
\usepackage{graphicx}
\usepackage{txfonts}
\usepackage{soul}
\usepackage{ulem}

\begin{document}

   \title{Interferometric imaging of the type~IIIb and U radio bursts observed with LOFAR on 22 August 2017}

  \author{Bartosz Dabrowski\inst{1}
          \and
          Katarzyna Mikuła\inst{1}
          \and
          Paweł Flisek\inst{1}
          \and
          Christian Vocks\inst{2}
          \and
          PeiJin Zhang\inst{3,4,5,6}
          \and
          Jasmina Magdaleni\'c\inst{7,8}
          \and
          Alexander Warmuth\inst{2}
          \and
          Diana E. Morosan\inst{9}
          \and
          Adam Froń\inst{1}
          \and
          Richard A. Fallows\inst{4}
          \and
          Mario M. Bisi\inst{10}
          \and
          Andrzej Krankowski\inst{1}
          \and
          Gottfried Mann\inst{2}
          \and
          Leszek Błaszkiewicz\inst{1}
          \and
          Eoin P. Carley\inst{5}
          \and
          Peter T. Gallagher\inst{5}
          \and
          Pietro Zucca\inst{4}
          \and
          Paweł Rudawy\inst{11}
          \and
          Marcin Hajduk\inst{1}
          \and
          Kacper Kotulak\inst{1}
          \and
          Tomasz Sidorowicz\inst{1}
          }

   \institute{Space Radio-Diagnostics Research Centre, University of Warmia and Mazury, R. Prawocheńskiego 9, 10-719 Olsztyn, Poland\\
              \email{bartosz.dabrowski@uwm.edu.pl}
        \and
    Leibniz-Institut f\"ur Astrophysik Potsdam (AIP), An der Sternwarte 16, 14482 Potsdam, Germany
        \and
    Institute of Astronomy and National Astronomical Observatory, Bulgarian Academy of Sciences, Sofia 1784, Bulgaria
        \and
    ASTRON~--~the Netherlands Institute for Radio Astronomy, Oude Hoogeveensedijk 4, 7991 PD Dwingeloo, the Netherlands
        \and
    Astronomy \& Astrophysics Section, Dublin Institute for Advanced Studies, Dublin 2, D02 XF86, Ireland
        \and
    CAS Key Laboratory of Geospace Environment, School of Earth and Space Sciences, University of Science and Technology of China, Hefei, Anhui 230026, China
        \and
    Solar--Terrestrial Centre of Excellence---SIDC, Royal Observatory of Belgium, 1180 Brussels, Belgium
        \and
    Centre for mathematical Plasma Astrophysics (CmPA), KU Leuven, Celestijnenlaan 200B, 3001 Leuven, Belgium
        \and
    Department of Physics, University of Helsinki, P.O. Box 64, Helsinki, Finland
        \and
    RAL Space, United Kingdom Research and Innovation (UKRI)~--~Science and Technology Facilities Council (STFC)~--~Rutherford Appleton Laboratory (RAL), Harwell Campus, Oxfordshire OX11 0QX, UK
        \and
    Astronomical Institute, University of Wrocław, Kopernika 11, 51-622 Wrocław, Poland\\
             }

   \date{Received 2021; accepted}

 
  \abstract
   {The Sun is the source of different types of radio bursts that are associated  with solar flares, for example. Among the most frequently observed phenomena are type~III solar bursts. Their radio images at low frequencies (below 100~MHz) are relatively poorly studied due to the limitations of legacy radio telescopes.}
   {We study the general characteristics of types~IIIb and U with stria structure solar radio bursts in the frequency range of 20\,--\,80~MHz, in particular the source size and evolution in different altitudes, as well as the velocity and energy of electron beams responsible for their generation.}
   {In this work types~IIIb and U with stria structure radio bursts are  analyzed using data from the  LOFAR telescope including dynamic spectra and imaging observations, as well as data taken in the X-ray range (GOES and RHESSI satellites) and in the extreme ultraviolet (SDO satellite).}
   {In this study we determined the source size limited by the actual shape of the contour at particular frequencies of type~IIIb and U solar bursts in a relatively wide frequency band from 20 to 80~MHz. Two of the bursts seem to appear at roughly the same place in the studied active region and their source sizes are similar. It is different in the case of another burst, which seems to be related to another part of the magnetic field structure in this active region. The velocities of the electron beams responsible for the generation of the three bursts studied here were also found to be different.}
   {}


   \maketitle
%

\section{Introduction}
   Type~III solar radio bursts are among the most frequently observed radio events, first described in a series of publications by Wild \& McCready \citep{wild1950, wild1950b, wild1950c}. They occur at frequencies in the range from  GHz down to kHz, with individual bursts lasting for several seconds, depending on the observing frequency. However, they sometimes appear in groups lasting from a few minutes   to over a day in the case of a storm \citep{benz2002}. The brightness temperature of these bursts usually lies between $10^{8}$ and $10^{12}$~K, sometimes reaching values of $10^{15}$~K \citep{dulk1985}. \cite{ginzburg1958} were the  first to present a type~III burst model based on plasma emission. Type~III bursts are believed to be the radio signature of electrons beams moving through the corona and into interplanetary space along open magnetic field lines \citep{lin1974}. The electron beams are generally considered to originate from magnetic reconnection or shocks \citep{dulk2000}. It is commonly believed that in type~III bursts, following their acceleration, faster electrons outpace the slower ones, resulting in a bump-on-tail instability in their velocity distribution. This leads to the generation of Langmuir plasma waves \citep{ginzburg1958}. The waves are later converted into radio waves at the local plasma frequency and its second or even higher harmonics \citep{bastian1998}. The plasma frequency depends on the electron density  \citep[e.g.,][]{warmuth2005}, which in general decreases with altitude in the solar corona. Therefore, the plasma frequency also decreases with height. 
   
   Type~III radio bursts are characterized by a high frequency drift rate. On the dynamic spectra, they resemble almost vertical structures. Their slope indicates fast moving electron beams accelerated in the corona. The drift rate on the metric wavelengths has been reported by various authors (see Table~\ref{table_drift}). It allows us  to estimate the velocity of the electrons beams responsible for the type~III bursts \citep[e.g.,][]{mann1999}. On the other hand, the electron density can be obtained from the  Newkirk (\citeyear{newkirk1961}) radial electron density model of the solar corona (hereafter called the  Newkirk model).
   
   The standard velocities of type~III radio bursts are about 0.1c and 0.6c \citep{benz2002}, but \cite{Melnik2021} observed this burst type  with URAN-2 and GURT in the frequency band 8\,--\,80~MHz, and showed that the velocities are in the range 0.17\,--\,0.2c (harmonic emission was assumed). On the other hand, \cite{morosan2014} observed type~III bursts with LOFAR in the 30\,--\,90~MHz band and obtained velocities in the range of 0.09\,--\,0.2c (harmonic emission for all the type III~bursts was assumed).

\begin{table}
\caption{Drift rates for type~III solar radio~bursts.}
\label{table_drift}
\centering
\begin{tabular}{lll}
\hline
\noalign{\smallskip}
Drift rate  & Freq. Range    & Reference\\
$\textsc{[MHz/s]}$      & $\textsc{[MHz]}$   &\\
\noalign{\smallskip}
\hline
\noalign{\smallskip}
$-0.01f^{1.84}$         & 0.075\,--\,550  & \cite{alvarez1973}\\
$-0.007354f^{1.76}$     & 0.04\,--\,85    & \cite{mann1999}\\
$-0.0672f^{1.23}$       & 10\,--\,80      & \cite{zhang2018}\\
\noalign{\smallskip}
\hline
\end{tabular}
\end{table}

   The variation in the source sizes of type III~bursts with frequency depends on such several effects, such  as (1) divergence of the magnetic field lines as a function of height in the corona; (2)   spatial width of the electron beam; (3) refractive propagation effects; and (4) scattering \citep[e.g.,][]{sainthilaire2013, zhang2021}. A study of source sizes with the Nançay Radioheliograph in the frequency range of $\sim$150\,--\,450~MHz performed by \cite{sainthilaire2013} shows that they are smaller at higher frequencies than at lower ones,  and that  their sizes decrease as $f^{-3.3}$. On the other hand, \cite{dulk1980} found that the sizes of the source at the frequencies of 80~MHz and 43~MHz are 11~arcmin and 20~arcmin, respectively. 
   Sometimes  type~III bursts show fine spectral substructures. In particular,  bursts classified as type IIIb are characterized by many narrowband bursts with slow drift, known as  stria bursts \citep{deLaNoe1972, abranin1979, mugundhan2017, sharykin2018, Melnik2018}. The bandwidth of the striae in the  20\,--\,80~MHz range is usually about 20\,--\,100~kHz \citep{deLaNoe1975, sharykin2018, chen2018} and the stria drift rates vary from 0 to 0.3~MHz/s \citep{sharykin2018}. The duration of the separate striae is short, typically around 1~s \citep{deLaNoe1975, chen2018}. The most typical scenario for the formation of striae is plasma density fluctuations along the electron beam path \citep{Takakura1975}. These small-amplitude perturbations can modulate the generation of Langmuir waves significantly \citep{Kontar2001}, and hence produce fine stria-like structures in the dynamic radio spectra.
   
   Bursts classified as type~U were first reported by \cite{Maxwell1958}. They got their name thanks to their shape in the dynamic spectrum, which resembles the inverted letter~U. They constitute a variant of type~III bursts;  they are a manifestation of the signatures of electron beams traveling along closed magnetic loops. Type U bursts are much rarer than type~III bursts. The rising and the descending branches of the type~U bursts have spatially separated sources \citep{Aurass1997}. Their polarization is generally low, below 10\%. In the meter  waves they were studied by \cite{Aurass1997}, \cite{Dorovskyy2015}, and \cite{Reid2017}.

   Several papers have recently been published, estimating the emission source sizes of type~IIIb bursts observed with the LOFAR telescope. \cite{kontar2017} obtained 400~arcmin$^2$ for 32.5~MHz. Its source was located at $x\approx280\textrm{ arcsec}$ and $y\approx390\textrm{ arcsec}$ from the solar  center. On the other hand, \cite{sharykin2018} found that source size for single stria equals 320~arcmin$^2$ (located around 290~arcsec, 320~arcsec) and 145~arcmin$^2$ (around $-30$~arcsec, 495~arcsec) for 30.11 and 41.77~MHz, respectively. It should be noted that in both publications, the authors obtained their results with the  tied-array beam forming method. In turn, the source size obtained by \cite{zhang2020} is about 100~arcmin$^2$ (its location was $-333$~arcsec, $-165$~arcsec from the solar center) for 26.56~MHz, and by \cite{murphy2021} is  150.6~arcmin$^2$ ($-1312$~arcsec, $-1064$~arcsec) for 34.76~MHz. In both papers the interferometric observation method was used. It is also worth mentioning  that \cite{dulk1980} estimated the effective source sizes (full widths at $1/e$ brightness) of the type~III bursts observed at 43~MHz around 314 arcmin$^2$. On the other hand, \cite{Chen2020} investigated how the scattering of radio waves on random density fluctuations affects the time profile evolution, sizes, and positions of the observed type III–IIIb solar radio bursts emitted via the plasma emission mechanism. They found that at 32.5~MHz the simulated source size for the fundamental emission ranged from around 60 to 100~arcmin$^2$ during the 2.5~s of the decay phase, which they found to be much smaller than typical observed source sizes from $\sim$300~arcmin$^2$ to $\sim$440~arcmin$^2$.
   
   In this article we study the general characteristics of two type~IIIb solar radio bursts and one U~burst with stria structures present (typical of type~IIIb bursts) observed on 22~August 2017 with the LOw-Frequency ARray (LOFAR; \citealt{vanhaarlem2013}) telescope in the frequency range of 20\,--\,80~MHz, including dynamic spectra and imaging observations. This gives the spatial information of radio bursts and relates it to the height, size, velocity, and energy of the electron beams responsible for their generation. In order to better understand the physical properties and evolution of the solar radio bursts, we used the data in X-rays from the  GOES and RHESSI satellites and in the extreme ultraviolet (EUV) from the Solar Dynamics Observatory (SDO).
 
   \begin{figure*}
   \centering
   \includegraphics[width=\hsize]{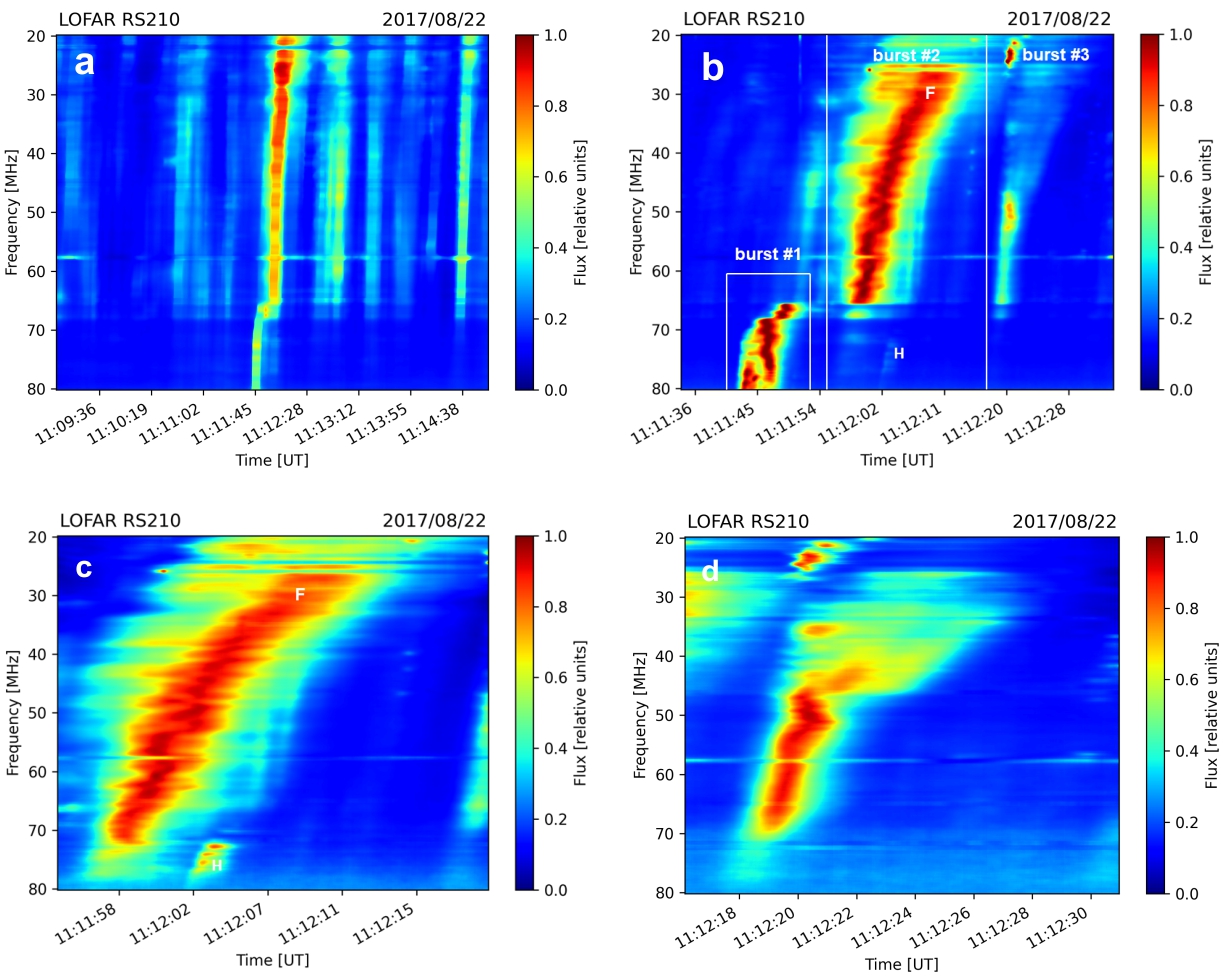}
      \caption{The solar dynamic spectra of the analyzed solar radio bursts.
 a) Part of the solar dynamic spectrum with the series of the type~III solar radio bursts recorded on 22~August 2017 between 09:00:00~UT and 13:59:59~UT by the LOFAR remote station RS210 in LBA band. b) Solar dynamic spectrum of the three investigated bursts, described as~\#1, \#2, and \#3. c) Solar dynamic spectrum of burst~\#2 with clearly visible fundamental (F) and harmonic (H) components. d) Solar dynamic spectrum of burst~\#3.}
   \label{dynamic_spectra}
   \end{figure*}

\section{Observations and data analysis}

   LOFAR is a large radio interferometer located in Europe, currently consisting of 52 stations, 38 of which  are located in the Netherlands and 14 in different countries across Europe: France, Germany, Ireland, Latvia, Poland, Sweden, and the United Kingdom. The 38 LOFAR stations in the Netherlands include the 24~core stations  (6 of which create the ``Superterp'') and 14 remote stations. For the whole International LOFAR Telescope (ILT) the largest baselines reach about 2000 kilometers. Each station of LOFAR observes with two types of antennas, the  Low Band Antennas (LBA), and the High Band Antennas (HBA), operating in the ranges 10\,--\,90~MHz and 110\,--\,240~MHz, respectively.
   
   Here we used part of the LOFAR spectral and imaging data from the observation project LC8\_013, ``Interferometric Observations of the Active Regions in Radio Domain Before and After the Total Solar Eclipse on 21 August 2017.'' For this research we incorporated 35 LOFAR stations (23 core and 12 remote stations), including the dynamic spectra and imaging observations at frequencies of 20\,--\,80~MHz (``LBA  Outer'' configurations). These 35 stations have a maximum baseline of about 58.4~km, which allows  a radio image with a theoretical spatial resolution of about 10.6~arcsec at 80~MHz \citep{vanhaarlem2013}. Unfortunately, due to coronal scattering, this resolution is limited to about 1~arcmin \citep{mercier2015}. We used Taurus~A as a calibrator to derive the stations’ gains. In order to obtain an interferometric image of the bursts we sampled the dynamic spectrum in 5~MHz intervals, which provided us with sufficient accuracy. The temporal and frequency resolutions of the dynamic spectra are 0.0104~s and 0.0154~MHz, respectively. The dynamic spectra were prepared   using the LOFAR Solar Imaging Pipeline developed at the Leibniz-Institut für Astrophysik Potsdam \citep{breitling2015}.
   
   The interferometric data were pre-processed with the Default Pre-Processing Pipeline (DPPP) \citep{vanDiepen2018}. The calibration of the gain solution is obtained from the simultaneous observation of the calibrator. The gain solution is then applied to the interferometry of the Sun to derive the calibrated amplitude and phase. After the pre-processing, the imaging is performed with WSClean \citep{offringa2014, offringa2017}. The integration time for each frame of imaging is 1.0~second, representing the time resolution of the imaging. After deconvolution and the CLEAN process performed with WSClean, we obtained the flux intensity distribution map in astronomical coordinates. Then we used the LOFAR-Sun toolbox \citep{zhang2020} to convert the coordinates and units and align them to solar observations made with other instruments (e.g.,~SDO).

   Using dynamic spectra it is possible to determine the parameters of type~III bursts, such as drift rates, frequency range, and  time span of each burst. In order to do this  computation, the raw LOFAR dynamic spectrum data were first cleaned using a median filter. For each burst we determined approximate burst boundaries in order to obtain a proper accuracy and diminish the risk of introducing other unassociated events into the analysis. We found the maximum intensity value in each frequency channel and established the time of its occurrence. Collected maxima determine the center-line of the analyzed burst. In order to obtain a smoother center-line, we used a moving average based in the frequency and time domains. A logarithmic function was fitted to the averaged maxima. The obtained values were used to calculate drift rates of the bursts with a 1~MHz interval. The error of the average drift rate is computed as the standard deviation of the drift rate at the frequencies that were determined earlier. On the basis of the obtained drift rates we determined the velocity and energy of electrons responsible for generation of the analyzed bursts \citep[e.g.,][]{dabrowski2021}. The reason for using the maxima of the bursts rather than their leading edges was the extremely irregular shape of the edges.

   In order to obtain an interferometric image of the burst we sampled the dynamic spectrum every 5~MHz. For each selected frequency channel we determined the maximum intensity of the burst and time of its occurrence. In the following step the contour of the radio emission area is set to half of the maximum intensity. Additionally, for each contour a centroid is calculated. The centroids of the bursts at each frequency were then used to estimate the height at which the bursts appeared and the velocity of the electrons beams responsible for their generation (according to the Newkirk model of electron density in solar corona). Moreover, it was possible to estimate the size of the radio emission area at a given height on the basis of the sizes of the contours. Furthermore, we used soft X-ray fluxes from GOES, hard X-ray spectroscopic and imaging data from RHESSI \citep{lin2002}, and observations in the extreme ultraviolet (EUV) from the Atmospheric Imaging Assembly (AIA) on board the Solar Dynamics Observatory (SDO) \citep{lemen2012}, using the 171~{\AA} channel (which corresponds to the quiet corona and upper transition region of the solar atmosphere).

   \begin{figure}
   \centering
   \includegraphics[width=\hsize]{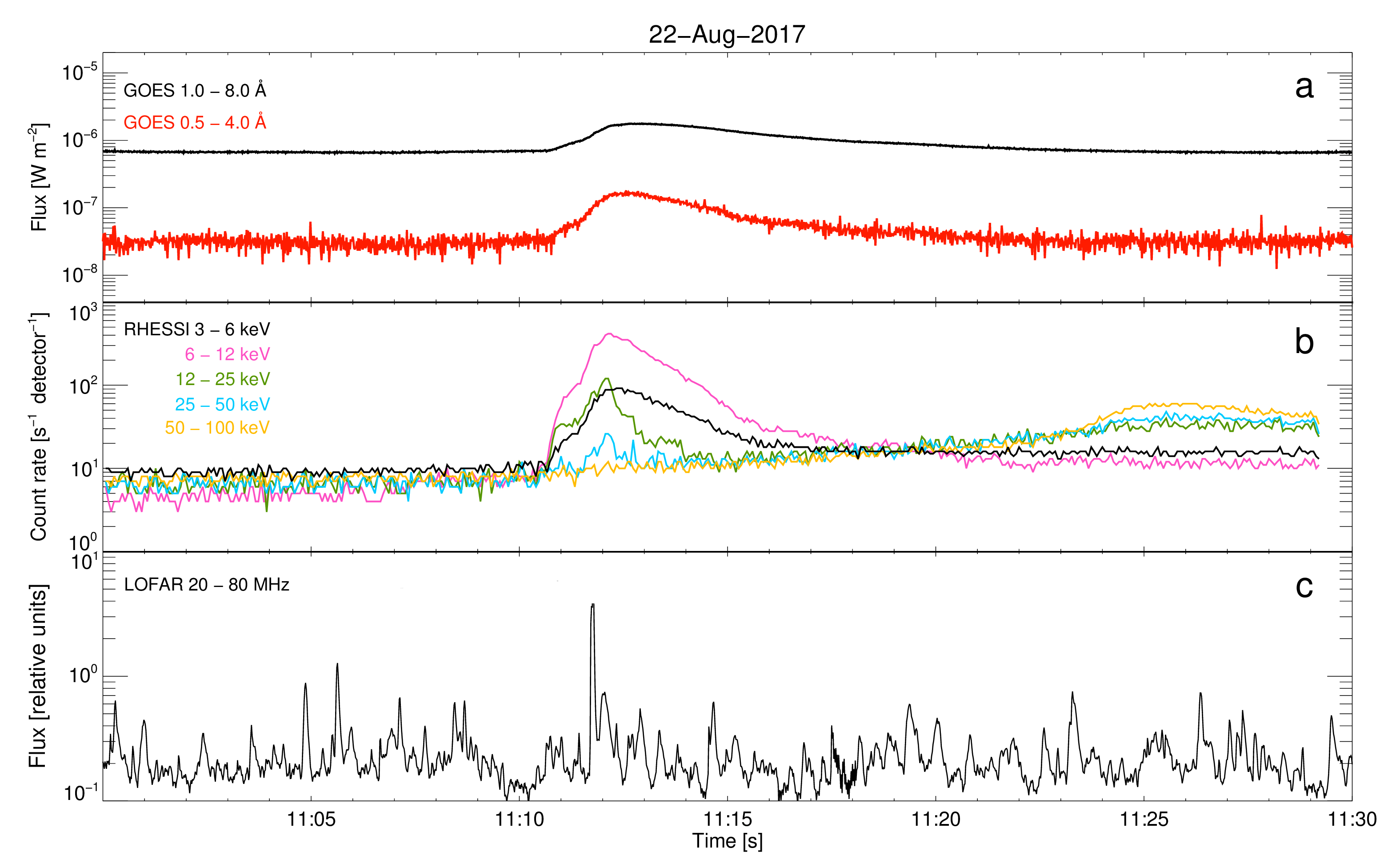}
      \caption{Comparison of radio flux with X-ray signals from GOES and RHESSI of the radio event observed on 22~August 2017. Panel (a) shows GOES fluxes, including a C1.0 solar flare with a peak at 11:13~UT. Panel (b) shows the RHESSI count rates in five wide energy bins. Panel (c) presents the radio total power flux from LOFAR 20\,--\,80~MHz (LBA) band with a single strong peak at 11:11:46~UT.}
   \label{signals_goes_rhessi_lofar}
   \end{figure}
   
   In our observations of the Sun we identified a type~III radio event observed with LOFAR on 22~August 2017 between 09:00:00~UT and 13:59:59~UT, which included a series of type~III solar radio bursts (Fig.~\ref{dynamic_spectra}a). We were unable to observe the beginning and end of the type~III storm because of the limited observing time. Particularly interesting  are the bursts that occurred around the single strong peak in the total radio flux at 11:11:46~UT, which appears as a rapid and short-lived increase by a factor of about~30 (but with no precursor activity), accompanied by a weak flare observed by GOES and RHESSI at about 11:13~UT (Fig.~\ref{signals_goes_rhessi_lofar}). Here we distinguished three solar radio bursts described as bursts~\#1, \#2, and \#3 (Fig.~\ref{dynamic_spectra}b). Burst~\#1 (type~U) occurred at 11:11:47~UT in the 65\,--\,80~MHz band; burst~\#2 (type~IIIb, Fig.~\ref{dynamic_spectra}c) was composed of two components, the fundamental (F), occurring at 11:12:00~UT in the 29\,--\,77~MHz band, and the  harmonic (H), which is weaker than the fundamental, occurring at 11:12:03 in the 73\,--\,79~MHz frequency band; and burst~\#3 (type~IIIb, Fig.~\ref{dynamic_spectra}d) at 11:12:20~UT in the  28\,--\,74~MHz band.
   
   The studied radio bursts were associated with a GOES~C1.0 solar flare originating from active region NOAA AR~12671 located at 5.73$^{\circ}$\,N, 25.84$^{\circ}$\,W. The flare started at 11:09~UT, reached a maximum at 11:13~UT, and ended at 11:16~UT. It was also registered by RHESSI in the hard X-ray range (Fig.~\ref{signals_goes_rhessi_lofar}) and observed in the EUV by the AIA in the 94~{\AA} and 171~{\AA} channels. The active region AR~12671 had a beta-gamma configuration of its photospheric magnetic field on 22~August 2017, and on that day was a source of many B and C~class flares. The sudden brightening observed at the western edge of the NOAA AR~12671, 4.58$^{\circ}$\,N, 32.6$^{\circ}$\,W is consistent with X-ray fluxes registered by GOES and RHESSI.

   \begin{figure*}
   \centering
   \includegraphics[width=\hsize]{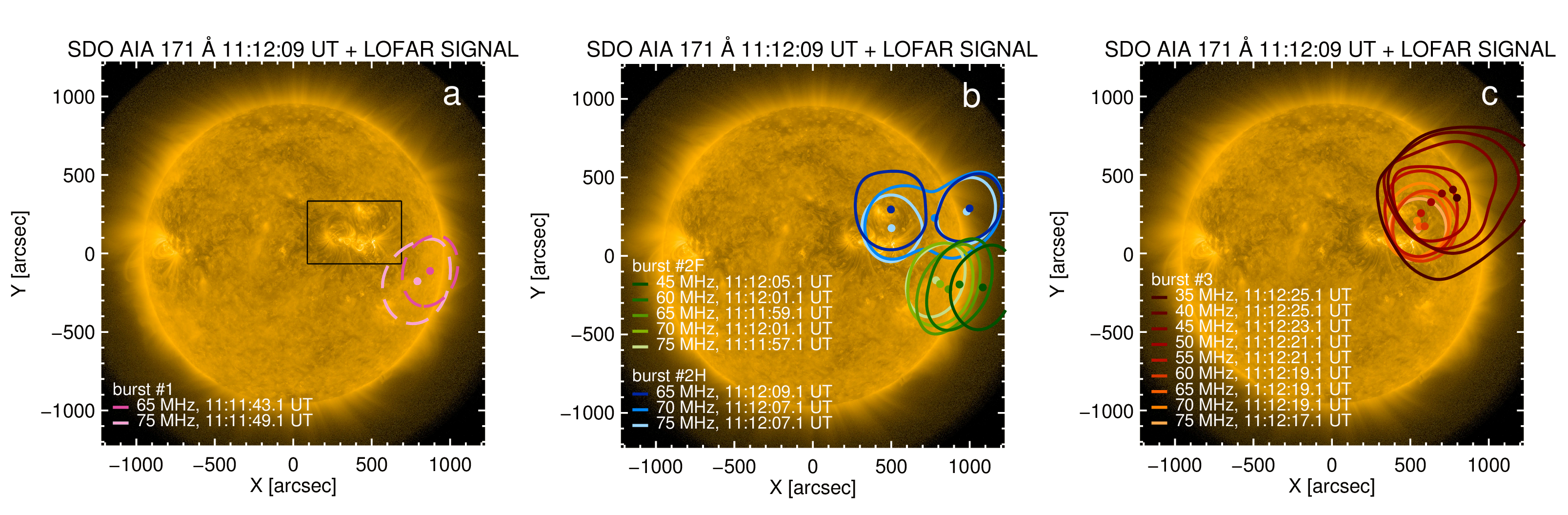}
   \caption{Image of the Sun received in the AIA~171~{\AA} channel by SDO with superimposed color contours showing bursts~\#1~(a), \#2~(b), and  \#3~(c) at a range of frequencies. In (a) the active region AR~12671 is indicated by a rectangle.}
   \label{uv_radio_b1_b2_b3}
   \end{figure*}

\begin{table*}
\caption{Physical parameters of bursts~\#1, \#2, and \#3 determined from the dynamic spectra and the centroids. Columns F and H are respectively the fundamental and harmonic components of  burst~\#2. The table also includes velocities determined on the basis of the drift rate obtained by various authors (see Table~\ref{table_drift}).}    
\label{table_parameters}
\centering
\begin{tabular}{lcccc}
\hline
\noalign{\smallskip}
                    & \multicolumn{1}{c}{Burst \#1}  & \multicolumn{2}{c}{Burst \#2}                       & \multicolumn{1}{c}{Burst \#3} \\
                    \noalign{\smallskip}
                    &                         & F                      & H                          &                                \\
\noalign{\smallskip}
\hline
\noalign{\smallskip}
\textbf{DYNAMIC SPECTRUM}    &                &                        &                            &                                \\
Time [UT]           & 11:11:47                & 11:12:00               & 11:12:03                   & 11:12:20                       \\
Freq. band [MHz]    & 80\,--\,65              & 77\,--\,29             & 79\,--\,73                 & 74\,--\,28                     \\
Height [R$_\odot$]  & 1.3\,--\,1.4            & 1.3\,--\,1.8           & 1.3\,--\,1.4               & 1.3\,--\,1.8                   \\
Central freq. [MHz] & 72.5                    & 53.1                   & 75.7                       & 50.8                           \\
Drift rate [MHz/s]  & $-2.6\pm0.2$            & $-4.3\pm2.1$           & $-4.9$\tablefootmark{*}    & $-9.0\pm4.8$                   \\
Energy [keV]        & $0.24\pm0.04$           & $1.76\pm1.73$          & 1.72                       & $9.06\pm9.92$                  \\
                    & \multicolumn{4}{c}{\textbf{Velocity [c]}}                                                                      \\ \cline{2-5}
\noalign{\smallskip}
This study          & $0.03\pm0.002$          & $0.08\pm0.04$          & 0.08                       & $0.19\pm0.10$                  \\
\cite{alvarez1973}  & 0.31                    & 0.29                   & 0.48                       & 0.28                           \\
\cite{mann1999}     & 0.16                    & 0.15                   & 0.25                       & 0.15                           \\
\cite{zhang2018}    & 0.15                    & 0.17                   & 0.23                       & 0.17                           \\
\noalign{\smallskip}
\hline
\noalign{\smallskip}
\textbf{CENTROIDS}  &                         &                           &                            &                               \\
Freq. band [MHz]    & 75\,--\,65              & 75\,--\,45                & --                         & 75\,--\,35                    \\
Height [R$_\odot$]  & 1.3\,--\,1.4            & 1.3\,--\,1.6              & --                         & 1.3\,--\,1.7                  \\
Velocity [c]        & --                      & $0.06\pm1.08\cdot10^{-4}$ & --                         & $0.09\pm3.2\cdot10^{-5}$      \\
Energy [keV]        & --                      & $0.92\pm0.003$            & --                         & $2.26\pm0.002$                \\
\noalign{\smallskip}
\hline
\end{tabular}
\tablefoot{
\tablefoottext{*}{We were able to estimate only two measuring points for the harmonic burst, which makes the drift rate error unobtainable.}\\
}
\end{table*}


\begin{table}
\caption{Source sizes of   bursts~\#1, \#2, and \#3 at different heights.}
\label{table_size}
\centering
\begin{tabular}{lccc}
\hline
\noalign{\smallskip}
                            & Burst \#1       & Burst \#2       & Burst \#3  \\
\noalign{\smallskip}
\hline
\noalign{\smallskip}
\begin{tabular}[c]{@{}l@{}}Height\\ {[R$_\odot$]}\end{tabular} & \begin{tabular}[c]{@{}c@{}}Source size\\ {[arcmin$^2$]}\end{tabular} & \begin{tabular}[c]{@{}c@{}}Source size\\ {[arcmin$^2$]}\end{tabular} & \begin{tabular}[c]{@{}c@{}}Source size\\ {[arcmin$^2$]}\end{tabular} \\
\noalign{\smallskip}
\hline
\noalign{\smallskip}
1.69 (35 MHz)               & --                     & --               & $205.9\pm1.3$    \\
1.61 (40 MHz)               & --                     & --               & $132.9\pm1.2$    \\
1.56 (45 MHz)               & --                     & $46.2\pm0.9$     & $102.4\pm1.3$    \\
1.51 (50 MHz)               & --                     & --               & $53.1\pm1.2$     \\
1.46 (55 MHz)               & --                     & --               & $56.2\pm1.3$     \\
1.43 (60 MHz)               & --                     & $50.2\pm1.0$     & $39.9\pm1.7$     \\
1.39 (65 MHz)               & $33.9\pm1.6$           & $46.7\pm1.7$     & $33.3\pm2.0$     \\
1.37 (70 MHz)               & --                     & $48.3\pm5.8$     & $42.8\pm2.6$     \\
1.34 (75 MHz)               & $49.6\pm11.8$          & $36.9\pm2.2$     & $30.9\pm5.6$     \\     
\noalign{\smallskip}
\hline
\end{tabular}
\end{table}

   \begin{figure}
   \centering
   \includegraphics[width=\hsize]{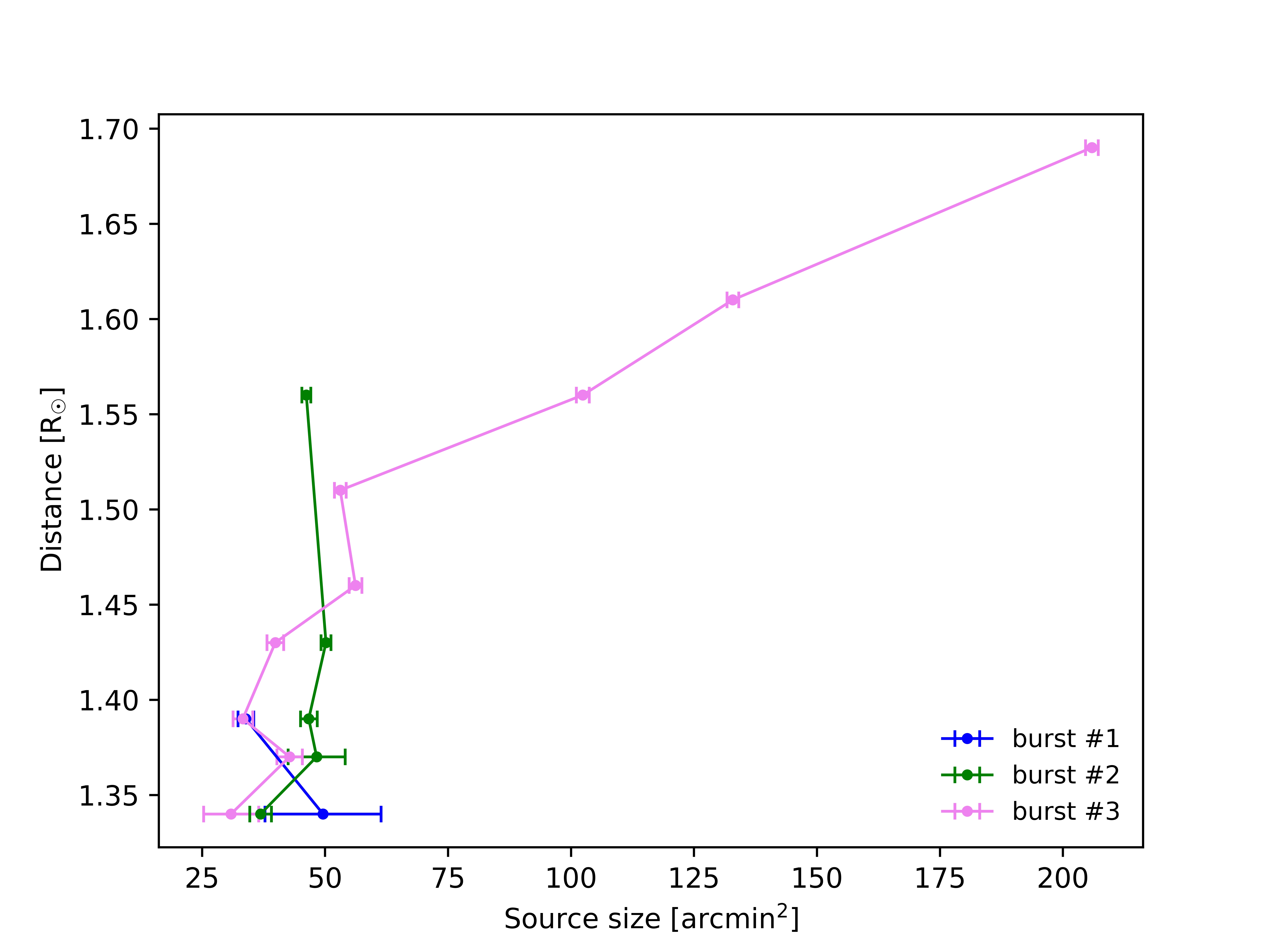}
      \caption{Change in the size of the emission area with the height for the investigated bursts (see Table.~\ref{table_size}). The horizontal error bars represent the beam size at  FWHM, which varies with frequency.}
   \label{burst_b1_b2_b3_size}
   \end{figure}

   \begin{figure*}
   \centering
   \includegraphics[width=\hsize]{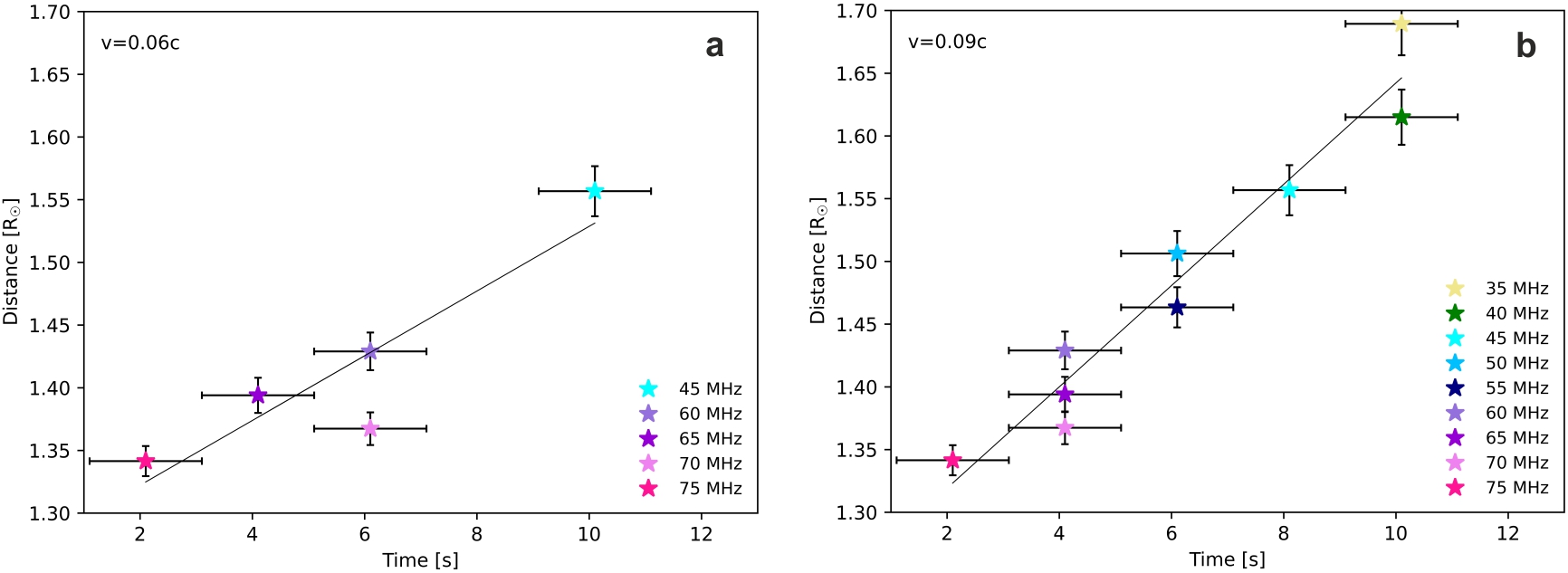}
   \caption{Radial motion of bursts~\#2 (left panel) and \#3 (right panel) obtained from centroids. The  colors correspond to different frequencies from 35 to 75~MHz. The velocities were calculated from the slopes of the fits and are given in the upper left corners. The vertical error bars in the two plots represent the extent of the FWHM of the beams, which  varies with frequency. The horizontal error bars are due to averaging the data in time steps of 1~s.}
   \label{b2_and_b3_motion}
   \end{figure*}

   \begin{figure*}[t]
   \centering
   \includegraphics[width=\hsize]{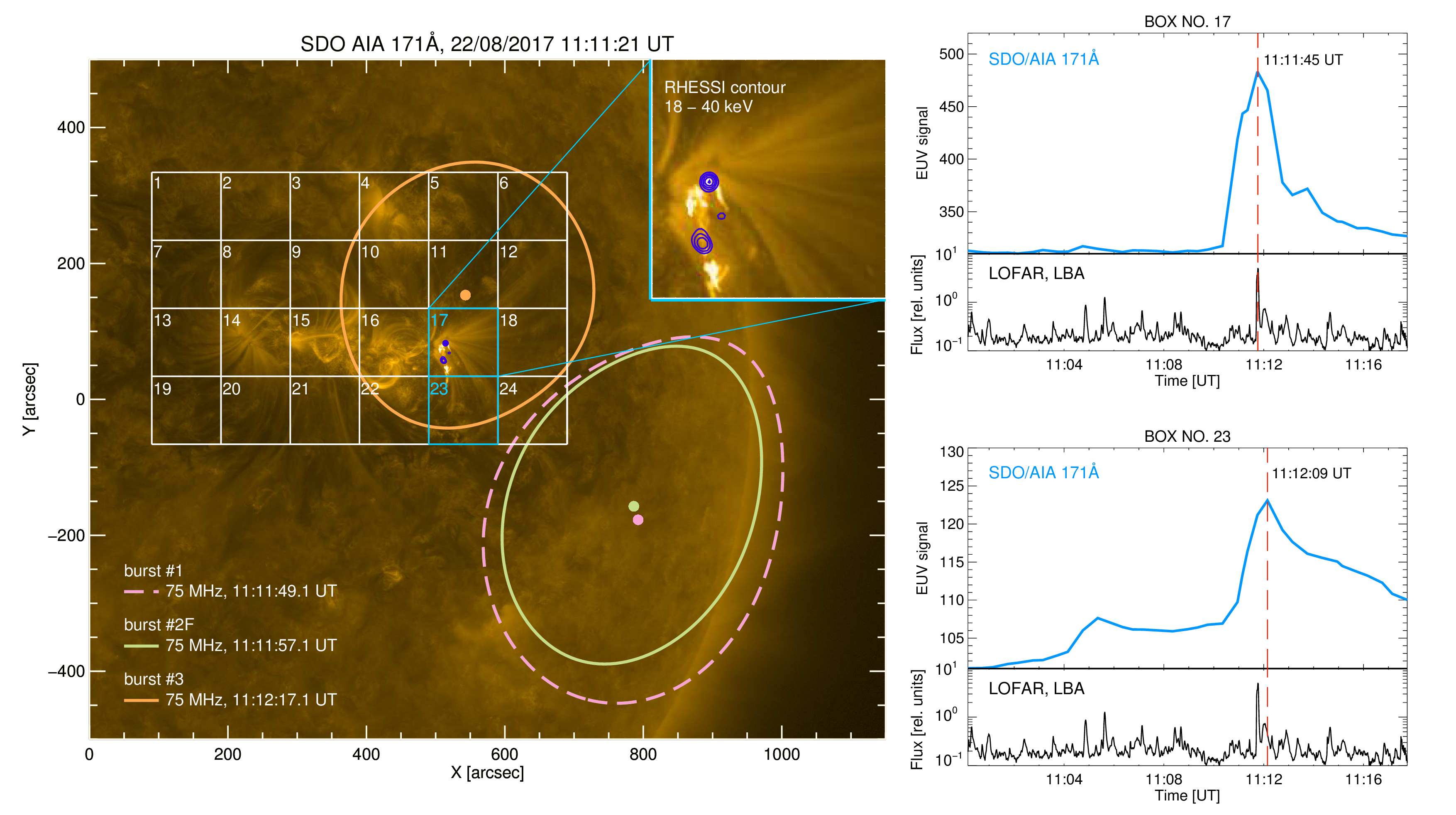}
   \caption{Map from SDO of the analyzed active region, and obtained signals from LOFAR and SDO. The left panel shows the active region AR~12671 during the solar flare registered in AIA~171~{\AA} by SDO with a superimposed grid and the radio contours at 75~MHz from the radio bursts~\#1, \#2, and \#3. In the right upper corner of the figure there is a magnification of box 17 with non-thermal hard X-ray footpoints (dark blue contour) in the 18~--~40~keV range, as derived from RHESSI observations. The right panels show EUV signals from boxes 17 and 23 and their comparison with LOFAR total radio flux. The maxima of the EUV signal are indicated by the red dotted lines.}
   \label{uv_grid}
   \end{figure*}
  
   \section{Results}
  
   The analyzed type~III radio bursts are outlined in Table~\ref{table_parameters}, which contains their physical parameters determined from the dynamic spectra and the centroids, as well as velocities determined on the basis of the drift rate obtained by various authors. Burst~\#1 was connected with the maximum increase in radio emission (Fig.~\ref{signals_goes_rhessi_lofar}c). Its time-flux profiles look peculiar. In the 65\,--\,75~MHz band there is a simple profile with a single very clear maximum peak. Above 75~MHz the time-flux profiles appear as double local maxima very close to each other. These time-flux profiles suggest that it may be a type~U burst where   stria structures (typical for type IIIb~bursts) are present. This type of burst is confirmed by the fact that the source region shows the usual bipolar structure associated with a flaring loop (which is discussed in detail below). Burst \#2 has   two components, fundamental (F) and harmonic (H). The mean ratio between them is about 1.7 for a given time (Fig.~\ref{dynamic_spectra}c). This is consistent with other observational values of 1.5\,--\,1.8 \citep{Suzuki1985}. The F~component of burst~\#2 at half the frequency of the H~component (around 35\,--\,40~MHz) is registered a few seconds later. There is a possibility that the harmonic emission is affected by the active region (Fig.~\ref{uv_radio_b1_b2_b3}). The explanation is that  the conditions for producing harmonic emission are not met immediately, but some time later. The type~IIIb burst~\#3 has a very peculiar kinked structure.
   
   The single strong peak, observed in the total radio flux at 11:11:46~UT, was connected with burst~\#1 and accompanied by a C1.0 flare (Fig.~\ref{signals_goes_rhessi_lofar}). A comparison of  the radio flux with the X-ray flux (in different energy ranges) shows that the maximum peak in the radio range was observed 18 to 60 seconds earlier than in the X-ray (Fig.~\ref{signals_goes_rhessi_lofar}). It is worth noting that in the energy range 12\,--\,25~keV, which is relevant for type~III electrons, we observe a jump (at 11:10:52~UT) to an enhanced level about 54 seconds before the radio flux peak (at 11:11:46~UT), although a maximum is reached later. This agrees with the general flare model that X-ray emission occurs before radio emission from the upper corona \citep[e.g.,][]{benz2017}.
   
   For the three analyzed bursts we received the radio images. In Fig.~\ref{uv_radio_b1_b2_b3} they are depicted as contours with their centroids superimposed on the image of the Sun obtained in the AIA~171~{\AA} channel by SDO. In the case of   burst~\#1, observed in the band 65\,--\,80~MHz, the radio images were taken only for 65 and 75~MHz just before the maximum phase in the given frequency because the maximum phase of the solar radio burst is too bright for the remote baseline in the observation of the calibrator to form interference correlation, which makes it impossible to calibrate the corresponding solar observation. We were also unable to obtain radio images for 70 and 80~MHz, due to radio interference. In the case of   burst~\#2,  observed in the band around 29\,--\,79~MHz we received images for 45, 60, 65, 70, and 75~MHz. It was not possible to obtain radio images for 30, 35, 40, 50, 55, and 80~MHz. This is due to the fact that in the lower and upper range of the observed radio band the relative intensity of this burst is quite low and its structure is irregular. In addition, there is radio interference at low frequencies. The radio images for  burst~\#3 (observed in 28\,--\,74~MHz band) were found from 35 up to 75~MHz. It was not possible to obtain a radio image for 30~MHz because in the lower range of the observed radio band the relative intensity of this burst is quite low and its structure is irregular.
   
   From the radio images we found the sizes of the radio emission sources for  three of the  analyzed bursts (Fig.~\ref{burst_b1_b2_b3_size} and Table~\ref{table_size}). For  burst~\#1 they could be determined only for two altitudes of 1.34~R$_\odot$ (75~MHz) and 1.39~R$_\odot$ (65~MHz), corresponding to the values of 49.6~arcmin$^2$ and 33.9~arcmin$^2$, respectively. In the case of burst~\#2 the sizes of the radio emission sources at different altitudes are similar, ranging from 36.9~arcmin$^2$ (for a height of 1.34~R$_\odot$) to 50.2~arcmin$^2$ for a height of 1.43~R$_\odot$. Burst~\#3 is a nice example of burst evolution through the corona. From the radio images we estimate that the burst starts at a distance of 1.34~R$_\odot$ (75~MHz) and ends at 1.69~R$_\odot$ (35~MHz). The source of the burst   moves outward   with decreasing frequency, as    expected with increasing coronal height, and gets larger. We observe a significant increase in the size of the source of the emission between the heights of 1.51~R$_\odot$ and 1.69~R$_\odot$, where it was 53.1~arcmin$^2$ and 205.9~arcmin$^2$, respectively. Here, at a distance of 0.18~R$_\odot$, the size of the source increases nearly four times, while below 1.51~R$_\odot$ the size of the source did not change that significantly. Between 1.34~R$_\odot$ and 1.51~R$_\odot$, corresponding to a distance of 0.17~R$_\odot$, the size of the source changed from 30.9~arcmin$^2$ to 53.1~arcmin$^2$. The analysis shows that for a fixed height of 1.34~R$_\odot$ (75~MHz) the size of the emission area equals 49.6, 36.9, and 30.9~arcmin$^2$ for burst~\#1, \#2, and \#3, respectively. 
   
   Determining   the velocities of the electron beams responsible for the emission of the investigated radio bursts and their energy was performed in two ways. The first used the dynamic spectra and the second   used the centroids of the emission sources obtained from interferometric radio maps.
   
   Assuming the Newkirk coronal density model, we found from the dynamic spectra that the velocity of the electron beams responsible for generating   the three analyzed bursts was between 0.03c and 0.19c. This corresponds to an energy of 0.24~keV and  9.06~keV, respectively (see Table~\ref{table_parameters}). On the other hand, the velocities calculated based on the slope of the fit were estimated for the centroid points for  burst~\#2 (only for the F~component) to be equal to 0.06c and for  burst~\#3 0.09c (Fig.~\ref{b2_and_b3_motion}). This corresponds to an energy of 0.92~keV and 2.26~keV, respectively (see Table~\ref{table_parameters}). It should also be noted that  the burst~\#3 velocity is significantly different from that  obtained from the dynamic spectrum because of the large projection effect visible in Fig.~\ref{uv_radio_b1_b2_b3}c. Unfortunately, it was not possible to determine the velocity from centroids for  burst~\#1 (because radio images were determined only for 65~MHz and 75~MHz) and for the H~component of burst~\#2 (due to the rather complex structure of the source).

   To identify which events in the active region AR~12671 were responsible for the observed radio bursts, we use observations of the solar flare registered on the SDO AIA~171~{\AA} image. On Fig.~\ref{uv_grid} we superimposed a grid of 24~boxes, each $100\times100$~arcsec in size. Then we calculated the signal from each box, which was summed, and in the next step divided by the exposure time of each SDO image. The obtained signals were enhanced in  boxes 17 and 23, where the C1.0 class solar flare occurred. The comparison of the EUV signal with the total radio flux in the frequency range 20\,--\,80~MHz shows a good relationship. The identification of the source location is independently confirmed by RHESSI X-ray imaging observations, which show a pair of non-thermal footpoints in the flaring area  seen by AIA (see hard X-ray footpoints in the
18~--~40~keV range in the  upper right corner of  Fig.~\ref{uv_grid}).
   
   Burst~\#1 is likely connected with emission from boxes 17 and 23, the maxima of which took place at 11:11:45~UT and 11:12:09~UT, respectively. On the other hand, the maximum of the EUV emission in box 23 appears 22~s later in comparison with the radio flux (Fig.~\ref{uv_grid}). At the beginning, the solar flare was observed in box 17 and then the plasma   moved along the magnetic field lines to the analyzed box 23. It should be noted that the RHESSI observations confirm that  burst~\#1 may be a type~U burst (Fig.~\ref{uv_grid}). Nevertheless, the analysis of its velocity was carried out in the same way as for bursts~\#2 and \#3. Burst~\#2 is probably connected with emission from boxes 17 and 23 (Fig.~\ref{uv_grid}). The propagation path of  burst~\#3 seems to be more radial than the other two bursts, indicating that the electron beam is propagating along a different set of magnetic field lines. The propagation direction of the beam is clearly shifted toward Earth, making the projection effects stronger. It was not possible to determine the exact location of  burst~\#3 due to the complex structure of this active region and numerous phenomena occurring in it (Fig.~\ref{uv_grid}). Based on the plasma motions along the magnetic field lines we can assume the burst is also associated with the solar flare. We have to take into account the temporal resolution of the SDO observations (around 24\,--\,36~s), which is lower than that of LOFAR, here reduced to 1~s.

\section{Discussion and conclusions}

   In this work we presented a study of the general characteristics of two type~IIIb solar radio bursts and one U~burst with stria structures present (typical for type~IIIb bursts), the most interesting among a series of bursts observed on 22~August 2017 with LOFAR in the frequency range of 20\,--\,80~MHz, including dynamic spectra and imaging observations. They are particularly interesting because they occurred around the maximum emission of this event in the radio range. Additionally, we also inspected the data recorded by instruments on board   GOES and  RHESSI (in X-ray) and SDO (in EUV), which complement the observations in the radio field.
   
   In this study we determined the source size limited by the actual shape of the contour at particular frequencies of type~IIIb and U solar bursts in a relatively wide frequency band from 20 to 80~MHz. The works by other authors \citep{kontar2017, sharykin2018, zhang2020, murphy2021} have focused on the determination of the emission size at a frequency of around 30~MHz with a 2D elliptical Gaussian function fit, which is only an approximation of their size \citep[e.g.,][]{kontar2019}. Therefore, comparison of our results with those obtained by other authors is not straightforward. In the analyzed case it was possible only for the 35 and 40~MHz frequencies and in our work, using the same method, we obtained 205.9~arcmin$^2$ and 132.9~arcmin$^2$ (both for  burst~\#3), respectively. At 35~MHz the source size we observed was about 1.4 times bigger than that obtained by \cite{murphy2021} (on the same frequency); both results were obtained using the same observation method (interferometric mode). A comparison of our results with those obtained by \cite{kontar2017} and \cite{sharykin2018} for the frequency of 32.5 and 30.11~MHz (and therefore very similar to ours) with the tied-array beam forming method shows that the source sizes we obtained with the interferometric method are 1.9 and 1.6 times smaller, respectively. In turn, for 40~MHz it has a similar value to those obtained by \cite{sharykin2018} (for 41.77~MHz), despite the different observation methods (tied-array beam and interferometric in our results). \cite{dulk1980} estimated that the effective source sizes of the type~III bursts observed at 43~MHz was about 314~arcmin$^2$. It should be noted that the source size was obtained with the use of the Culgoora radioheliograph. At 45~MHz, the source size we observed was about three and seven times smaller (for burst~\#3 and \#2, respectively) than that obtained by \cite{dulk1980}.
   
   In the case of burst~\#3 the size of the emission source increases with altitude (Fig.~\ref{burst_b1_b2_b3_size}). This is well-known and can be explained by coronal scattering of radio waves and the cross-sectional size of the electron beam as it moves along magnetic field lines \citep{zhang2020}. For burst~\#2 this is different; the source sizes at different altitudes are similar. Since a big difference in scattering is not to be expected, which suggests that the magnetic field configuration is less divergent than for burst~\#3.
   
   The reason why it was difficult to compare the obtained emission source sizes of the analyzed bursts with the results received by other authors was the lack of observations carried out at low frequencies with sufficient spatial resolution. On the other hand, in several studies (\citealt{kontar2019} and references therein), the angular source sizes of type~III bursts were determined. On that basis, \cite{kontar2019} found the relation of the angular source size of type~III emission (determined on the FWHM) versus frequency in the band ~0.05 to 500~MHz. It turns out that the source size increases with the frequency and is described by the following equation: size$=(11.78\pm0.06)\cdot f^{-0.98\pm0.05}$ (where the source size is given in degrees and the frequency in MHz). Assuming that the source is circular, from this relation (which in some cases is far from the truth) we can estimate its size. In the case of the analyzed radio bursts observed in the 35\,--\,75~MHz band, the sizes of the emission areas are 369~arcmin$^2$ and 83~arcmin$^2$, respectively. Comparison of the source sizes obtained for the bursts analyzed here at 65 and 75~MHz (only at these two frequencies was it   possible to determine the sizes of the emission for all three bursts) with the sizes determined from the dependence described above  shows that at 65~MHz the size of the sources we obtained is from 2.3 (for burst~\#2) to approximately 3.2 (for bursts~\#1 and \#3) times smaller. Furthermore, for 75~MHz our source size is 1.7 to 2.7 times smaller (for all of the bursts).
   
   Coronal scattering might affect the apparent positions of the sources. According to the method proposed by \cite{chrysaphi2018} we estimate that the source observed at 35~MHz and 75~MHz (corresponding to the bandwidth in which  burst~\#3 was observed) to be radially shifted by about 0.6~\(\textup{R}_\odot\) and 0.2~\(\textup{R}_\odot\), respectively, from its true location. For this reason the  apparent positions of sources observed at 35~MHz and 75~MHz is $R_{35}=1.69+0.6=2.29~\textup{R}_\odot$ and $R_{75}=1.34+0.2=1.54~\textup{R}_\odot$. The sources at this position correspond to the corona described by the $4.3\times$ and $3.0\times$ Newkirk density model, respectively. The source observed at 35~MHz is shifted more than the source at 75~MHz. If the apparent sources are interpreted as true sources, an increase in density due to the radial shifting is often observed.
   
   Bursts~\#1 and \#2F (at 75~MHz, see   Fig.~\ref{uv_grid}) seem to appear at roughly the same place in active region AR~12671. The bursts are quite different. Burst~\#1 is a U-burst, so the electrons are injected into a closed loop, while burst~\#2 extends to lower frequencies, so the electrons in this burst have been injected into open magnetic field lines. This means that  a small variation in source location leads to completely different burst characteristics. It also turned out that the mean values of their source sizes are similar. It is different in the case of burst~\#3, which seems to be related to another part of the magnetic field structure in this active region.
   
   Estimated velocities (and energy) of bursts~\#1, \#2, and \#3 generally stick to the lower boundary of standard type~III radio burst velocities of 0.1\,--\,0.6c \citep{benz2002}. Burst~\#1 can be classified as a U~burst in which   stria structures are present. The low electron velocity (equal to 0.03c) of this burst is due to the magnetic field line having a significant horizontal component. Then the projection of the electron velocity on the radial direction, which determines the drift rate, can be quite low. This certainly applies to a U~burst as the magnetic loop along which the electrons are moving becomes horizontal at some point. The velocities of bursts~\#2 and \#3 are equal to 0.1 and around 0.1\,--\,0.2, respectively, and are comparable with the results obtained by \cite{morosan2014}, who received 0.09\,--\,0.2c (harmonic emission for all the type~III bursts was assumed) in the 30\,--\,90~MHz LOFAR band. We can conclude that the different values of velocity of the electron beam responsible for the generation of analyzed bursts  suggest their propagation in the solar corona with different geometric configuration of the magnetic field. We also observe different energies of the electron beams responsible for their formation. It ranges from around 0.2 to over 9.0~keV and is different for each burst.
  
   Comparison of radio and X-ray fluxes shows that the maximum peak in the radio range was observed up to about 1~min earlier than in the X-ray, but about 1~min before the maximum we also observed a jump in the energy range 12\,--\,25~keV. This agrees with the standard solar flare model.
  
   The radio images of the three bursts observed at different frequencies allowed the determination of their height in the solar corona and estimation of the size of their emission. In turn, the images in EUV from SDO (i.e., from the  AIA~171~{\AA} channel) gave us information about the processes taking place in the upper transition region. Thanks to imaging at three wavelengths (EUV, X-ray, and radio) it was possible to better understand the observed event.

\begin{acknowledgements}
   We acknowledge the National Science Centre, Poland and the Deutsche Forschungsgemeinschaft (DFG, German Research Foundation) for granting ``LOFAR observations of the solar corona during Parker Solar Probe perihelion passages'' in the Beethoven Classic~3 funding initiative under project numbers 2018/31/G/ST9/01341 and VO 2123/1-1, respectively. UWM would like to thank the Ministry of Education and Science of Poland for granting funds for the Polish contribution to the International LOFAR Telescope (decision number 2021/WK/2) and for maintenance of the LOFAR PL-612 Baldy (decision number 59/E-383/SPUB/SP/2019.1). Diana E.~Morosan acknowledges the Academy of Finland project `RadioCME' (grant number 333859). The authors thank Derek McKay (Aalto University Metsähovi Radio Observatory, Finland) and KSP team for support in solar LOFAR~research. This paper is based on data obtained with the International LOFAR Telescope (ILT) under project code LC8\_013. LOFAR \citep{vanhaarlem2013} is the Low Frequency Array designed and constructed by ASTRON. It has observing, data processing, and data storage facilities in several countries that are owned by various parties (each with their own funding sources), and that are collectively operated by the ILT foundation under a joint scientific policy. The ILT resources have benefited from the following recent major funding sources: CNRS-INSU, Observatoire de Paris and Université d’Orléans, France; BMBF, MIWF-NRW, MPG, Germany; Science Foundation Ireland (SFI), Department of Business, Enterprise and Innovation (DBEI), Ireland; NWO, The Netherlands; The Science and Technology Facilities Council, UK; Ministry of Science and Higher Education, Poland. Thanks to Nicolina Chrysaphi for sharing the codes (v2.1p) for estimation of scattering effects and Tomasz Mrozek (Space Research Centre, PAS, Solar Physics Division, Poland) for help with RHESSI software. We acknowledge the referee for comments and suggestions which helped us in bringing out the results more clearly.
\end{acknowledgements}


   \bibliographystyle{aa}
   \bibliography{dabrowski.bib}


\end{document}